\documentclass[10 pt]{article}

\usepackage[utf8]{inputenc} 
\usepackage[T1]{fontenc}    
\usepackage{hyperref}       
\usepackage{url}            
\usepackage{booktabs}       
\usepackage{amsfonts}       
\usepackage{amsmath}
\usepackage{nicefrac}       
\usepackage{microtype}      
\usepackage{graphicx}
\usepackage{spconf}
\usepackage{tabularx}
\usepackage{lipsum}
\usepackage{dsfont}
\usepackage{multirow}
\usepackage{caption}
\usepackage[shortlabels]{enumitem}

\newcommand\blfootnote[1]{%
  \begingroup
  \renewcommand\thefootnote{}\footnote{#1}%
  \addtocounter{footnote}{-1}%
  \endgroup
}

\title{Multimodal Fusion of Imaging and Genomics for Lung Cancer Recurrence Prediction}

\name{Vaishnavi Subramanian$^{1}$ \  Minh N. Do$^{1}$ \  Tanveer Syeda-Mahmood $^{2}$ }

\address{$^{1}$ Electrical and Computer Engineering, University of Illinois at Urbana-Champaign, USA\\
    $^{2}$ IBM Research, Almaden Research Center, San Jose, USA}

\begin{document}

\maketitle

\begin{abstract}
Lung cancer has a high rate of recurrence in early-stage patients. Predicting the post-surgical recurrence in lung cancer patients has traditionally been approached using single modality information of genomics or radiology images. We investigate the potential of multimodal fusion for this task. By combining computed tomography (CT) images and genomics, we demonstrate improved prediction of recurrence using linear Cox proportional hazards models with elastic net regularization. We work on a recent non-small cell lung cancer (NSCLC) radiogenomics dataset of 130 patients and observe an increase in concordance-index values of up to 10\%. Employing non-linear methods from the neural network literature, such as multi-layer perceptrons and visual-question answering fusion modules, did not improve performance consistently. This indicates the need for larger multimodal datasets and fusion techniques better adapted to this biological setting.
\end{abstract}

\section{Introduction}


Non-small cell lung cancer (NSCLC) is the most common among lung cancers, the world's leading cause of cancer deaths. 
Most NSCLC cancers are diagnosed at a late stage leading to  poor prognosis. However, even in cancers which are detected early, about 30\% - 55\% of patients develop recurrence despite curative resection~\cite{uramoto2014recurrence}. Recurrence is attributed to different causes, including underestimation of tumor stage and dissemination of cancer cells during surgery. 
\blfootnote{*This work was supported by the IBM-Illinois C3SR center and the Jump-ARCHES grant.}

Prediction of recurrence risk post-surgery has been attempted using 
individual modalities about the patients. Genomics, which captures the cellular properties of cancerous tissue, have been used through analysis of microRNA~\cite{lu2012microrna}
and gene expression data~\cite{lee2008prediction}. 
At the other end of the spectrum, radiology imaging such as computed tomography (CT) provides an overall picture of the patient's tumour at a larger physical scale. \textit{Radiomic} features derived from CT images have been extracted from regions of interest and used to predict recurrence~\cite{d2019ct,starkov2019use}.

With the increase in the availability of multimodal public datasets which provide patient information for more than one modality, such as TCGA-TCIA~\cite{weinstein2013cancer},
there is potential to greatly improve recurrence prediction by fusing information from imaging and genomics since they originate from different physical scales. 
This has been demonstrated recently for survival prediction with histology and genomics~\cite{yao2017deep,huang2018deep}. 

\begin{figure}[t]
    \centering
    \includegraphics[width=0.45\textwidth]{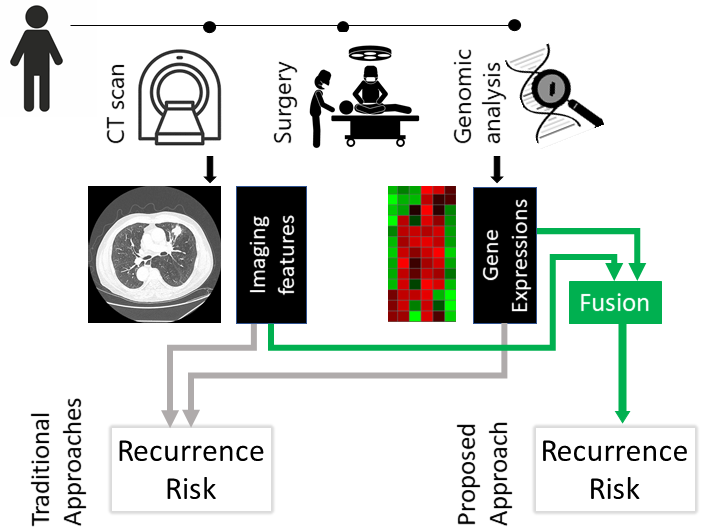}
    \caption{Overview: Traditional methods attempt to predict recurrence risk in patients from individual modalities of imaging and genomics. We propose to fuse the information to better predict recurrence risk.}
    \label{fig:pipelines}
\end{figure}

In our work, we perform multimodal fusion of CT images (radiology) and RNA-sequencing based gene expression data (genomics) for post-surgical recurrence prediction in NSCLC, as shown in Fig.~\ref{fig:pipelines}. We extract features from the two modalities and input these to different models. 

We first apply linear Cox proportional hazards model with regularization to understand the potential of multimodal fusion techniques. We then consider neural networks as models for the recurrence risk prediction task. Here, we consider multi-layer perceptrons and fusion methods proposed recently for visual question answering (VQA)~\cite{yu2018beyond,BenYounes_2019_AAAI}.
Our evaluation is based on concordance indices, which quantify the extent of correct ordering of risk scores, and time-dependent AUC curves, which measure the classification ability of the models. 

\section{Preliminaries}

\subsection{Data and feature extraction}

We work with the NSCLC-Radiogenomics dataset~\cite{bakr2018radiogenomic} of 211 patients. CT imaging, gene expression and recurrence information are all available for 130 of these patients. Further, those patients who underwent adjuvant/neo-adjuvant therapy in addition to surgical resection were removed from the patient cohort, yielding 107 patients. This is done to avoid any influence of non-surgical intervention mechanisms on cancer growth suppression. 

The gene expression (RNA-sequencing) data was available for all patients for 5268 genes. The top 500 most variant genes in terms of RNAseq expression were chosen to pick a subset of most relevant genes. 
This served directly as the genomics feature vector ($f_g$). 

The segmentations for nodules in the CT images  obtained from the dataset were processed using the DICOM Toolkit $\texttt{dcmtk}$ to obtain nodule masks. These masks were fed into a standard radiomics library~\cite{van2017computational} to obtain a 107-dimensional PyRadiomic (PyRad) features ($f_p$) capturing texture and intensity patterns of the nodule and its immediate surrounding tissues of the lung. In addition, the different 2D slices within the mask area were processed through an Imagenet-pretrained DenseNet~\cite{huang2017densely} to extract slice-level features, which were aggregated via maximum-pooling across slices to obtain a 1024-dimensional DenseNet (DN) feature ($f_d$) for each patient. 

Although features can be learnt from the data, we work with pre-computed features as the first step in exploring multimodal fusion for recurrence prediction due to the small-sized dataset.
A 5-fold cross validation was used on 70-10-20 splits for training, validation and testing with stratification based on the presence/absence of recurrence. Code and splits are available at \url{https://github.com/svaishnavi411/multimodal-recurrence}. 




\subsection{Survival analysis}
Survival analysis, similar to regression tasks, trains models to correctly assign a risk-value to each sample. However, unlike regression, it is not necessary that all samples have experienced the event of interest. Each sample $i$ has two associated values:
\begin{enumerate}[(i)]
\setlength{\itemsep}{0em}
    \item A binary event value $e_i$, which indicates whether the patient experienced a recurrence ($e_i =$ 1) or no recurrence ($e_i =$ 0) in the study's observation period. 
    \item An associated numeric value $t_i$ which equals the time duration of observation from surgery if there was no recurrence ($e_i = 0$), and the time to recurrence from surgery if there was a recurrence ($e_i = 1$). 
    That is, $t_i = ( e_i (\text{time-to-recurrence}) + (1-e_i) (\text{time-under-observation}) )$.
\end{enumerate}

%
\subsection{Evaluation metrics}
The concordance-index (C-index) is the most commonly used metric for survival analysis. It quantifies the effectiveness of a given risk-prediction algorithm in correctly ordering events. Consider values $\{e_i\}_{i=1}^N$, $\{t_i\}_{i=1}^N$ defined previously and let $\{o_i\}_{i=1}^N$ denote the predicted recurrence risk-score for all $N$ patients in the data. It is desired to predict higher recurrence risk scores $o_i$ for patients with a lower $t_i$ if the recurrence was observed, $e_i = 1$. This is captured in the C-index, defined as
$$ \text{C-index} = \frac{1}{n} \sum_{i \in \{{1 \dots N : e_i = 1}\}} \sum_{t_j > t_i} \mathds{1}[o_i > o_j],$$ where $n$ is the number of ordered pairs in the data. 
High C-index values are desired. A C-index of 0.5 is equivalent to a random guess. 

We also make use of the time-dependent AUC (TD-AUC) which is an extension of the traditional area under receiver operating characteristic (AUC).
For the binary classification problem of distinguishing patients who have experienced recurrence at time $t$ to those who have not, the TD-AUC measures the AUC as a funciton of time $t$. 
Both metrics are calculated using the \texttt{scikit-survival} package~\cite{polsterl2015fast}.


\begin{table*}[t]
\caption{Comparison of C-Index for linear and non-linear models across 5 folds, including genomics (G), pyradiomics (P), densenet (D), early fusion (EF), intermediate fusion (IF), late fusion (LF), block superdiagonal tensor fusion (BLOCK) and multimodal factorized higher-order pooling (MFH). The best C-index for each fold is shown in bold.}
\begin{minipage}{0.49\textwidth}
\caption*{(a) Linear models. }
\label{tab:linear}
\resizebox{\columnwidth}{!}{
\begin{tabular}{l|lllll}
\hline
Method & Fold 1 & Fold 2 & Fold 3 & Fold 4 & Fold 5 \\
\hline
Cox(G) & 0.52 & 0.45 & \textbf{0.60} & 0.72 & 0.75 \\
Cox(P) & 0.46 & 0.39 & 0.30 & 0.37 & 0.80\\
Cox(D) & \textbf{0.53} & 0.49 & 0.44 & 0.65 & 0.52 \\
\hline
Cox(EF(G, P)) & 0.41 & 0.36 & 0.46 & 0.57 & 0.81 \\
Cox(LF(G, P)) & 0.41 & 0.42 & 0.49 & 0.60 & 0.81 \\
Cox(EF(G, D)) & \textbf{0.53} & 0.46 & \textbf{0.60} & 0.80 & 0.77  \\
Cox(LF(G, D)) & 0.49 & \textbf{0.50} & 0.55 & \textbf{0.82} & 0.59 \\
\hline
Cox(EF(G, P, D)) & 0.43 & 0.37 & 0.47 & 0.61 & \textbf{0.82} \\
Cox(LF(G, P, D)) & 0.46 & 0.46 & 0.51 & 0.81 & 0.60 \\
\hline
\end{tabular}
}

\end{minipage}%
\begin{minipage}{0.02\textwidth}
\
\end{minipage}%
\begin{minipage}{0.47\textwidth}
\caption*{ (b) Non-linear models.
}
\label{tab:non-linear}
\resizebox{\columnwidth}{!}{
\begin{tabular}{l|lllll}
\hline
Method & Fold 1 & Fold 2 & Fold 3 & Fold 4 & Fold 5 \\
\hline
MLP(G) & 0.52 & 0.43 & 0.64 & 0.50 & 0.64  \\
MLP(P) & 0.55 & 0.43 & 0.28 & 0.68 & 0.47  \\
MLP(D) & 0.41 & \textbf{0.65} & 0.74 & 0.25 & 0.46 \\
\hline 
\hline
IF(G, P) & \textbf{0.60} & 0.42 & 0.32 & 0.72 & 0.63 \\
LF(G, P) & 0.51 & 0.57 & \textbf{0.75} & 0.42 & \textbf{0.83} \\
IF(G, D) & 0.50 & 0.50 & 0.50 & 0.50 & 0.50 \\
LF(G, D) & 0.46 & 0.50 & 0.32 & 0.46 & 0.29 \\
\hline
BLOCK(G, P) & 0.57 & 0.43 & 0.30 & 0.63 & 0.36 \\
MFH(G, P) & 0.50 & 0.50 & 0.50 & 0.50 & 0.50 \\
BLOCK(G, D) & \textbf{0.60} & 0.52 & 0.73 & \textbf{0.77} & 0.72 \\
MFH(G, D) & 0.56 & 0.52 & 0.51 & 0.34 & 0.76 \\
\hline
\end{tabular}
}
\end{minipage}
\end{table*}
\begin{figure*}[t]
    \centering
    \includegraphics[width=\textwidth]{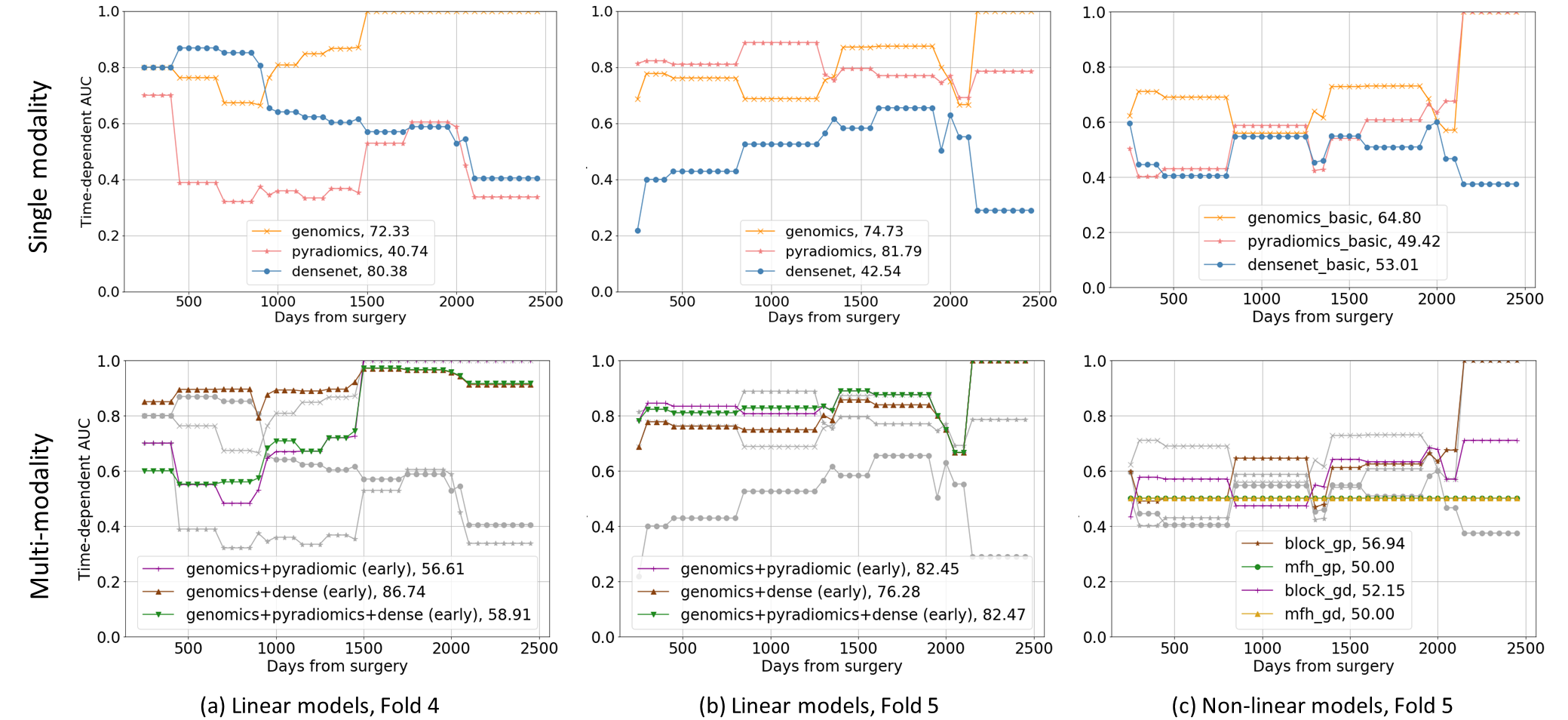}
    \caption{Time Depending AUCs. (a) Fold 4, linear (b) Fold 5, linear, and (c) Fold 5, non-linear. Top row: Single modality, bottom row: fusion models with single modality curves grayed out for comparison. Accompanying number quantifies the mean AUC.}
    \label{fig:time_varying_auc}
\end{figure*}

\section{Linear Methods}

\subsection{Cox proportional hazard model with regularization}

We first work with the Cox proportional hazards model~\cite{cox1972regression} which is a commonly used semi-parametric model in survival analysis. It works by estimating the hazard function $h(t)$, representing the risk of an event at time $t$, based on a linear combination $\beta$ of the feature weights $\mathbf{x}$ for a fixed patient
$$ h(t|\mathbf{x}) = h_0(t) \exp(\beta^T \mathbf{x}) ,$$
where $h_0(t)$ is a shared baseline hazard. 
The elastic net regularization is frequently imposed on the weights $\beta$ to allow more features than number of samples~\cite{friedman2010regularization}. The corresponding constraint is a weighted combination of L1 and L2 penalties:
$$\alpha\sum_i |\beta_i| + (1-\alpha)\sum_i\beta_i^2 \leq c.$$ 
An efficient algorithm to solve for $\beta$ under the given constrains was proposed in ~\cite{friedman2010regularization}. We use the algorithm's implementation from the \texttt{scikit-survival} package~\cite{polsterl2015fast}.

\subsection{Experiments and results}

We make use of the Cox proportional hazards model with the elastic net regularization (CoxNet) to predict recurrence. For each fold, we fit the model on the training set and report performance on the testing set. The validation set is not used here. 

To set the baseline performance of individual modalities, the 500-\textit{d}, 107-\textit{d} and 1024-\textit{d} features corresponding to genomics, pyradiomics and densenet are fed directly to the CoxNet. We work with two fusion techniques - (i) early fusion ($f_1, f_2, \dots f_n$), where a concatenation of all the features $f_1$, $f_2, \dots f_n$ are fed to the CoxNet, and (ii) late fusion ($f_1, f_2, \dots f_n$), where models corresponding to features $f_1, f_2  \dots f_n$ are trained and the output risks are aggregated. These are employed on combinations of ($f_g, f_p$), ($f_g, f_d$) and ($f_g, f_p, f_d$). 

%
The C-index values are reported in Table~\ref{tab:linear}(a). We make the following observations. For folds 1 through 3, the performance of the single modality models are poor, which result in a poor performance of the fusion methods as well. This poor performance could be attributed to the mismatched statistics of data between the training and testing sets for these folds, since the dataset is small. 

The C-index values for corresponding to $f_g$ and $f_d$ are moderately high for fold 4, while those of $f_g$ and $f_p$ are good for fold 5. The fusion of these modalities further improves the C-index by 10\% in fold 4 but only 2\% in fold 5. 

To further understand the utility of fusion, the time-dependent AUCs are plotted in Fig.~\ref{fig:time_varying_auc} for early fusion models for folds 4 and 5 in columns (a) and (b) respectively. In fold 4, the early fusion of $f_g$ and $f_d$ results in the model learning better weights across time which outperform the best single modality models. In fold 5, the model gets contrasting signals from different models and chooses a middle path for the time frame close to 1000 days from surgery. Across these folds, there does not exist one feature which consistently outperforms the rest. This demonstrates the potential of fusion. 

While folds 4 and 5 show promise, a linear fusion methodology does not allow for more complex connections to be explored between the different modality features, which could potentially result in better fusion models. Thus, we explore neural networks for fusion and prediction in the next section.
\section{Non-Linear Methods}

\subsection{Neural networks}
\label{nn_methods}
In contrast to Cox proportional models which work on linear combinations of the feature weights $\mathbf{x}$, non-linear methods allow more complex relations to be learnt. 
To set the non-linear baseline, we first train single modality multi-layer perceptrons (MLPs) which take as input the corresponding single modality feature and return the recurrence-risk for each patient. 
We then investigate four different methods of fusion. 
Two of these are multi-stream MLPs:
intermediate fusion MLP (IF) 
\begin{align}
\setlength{\itemsep}{0em}
    &f_g' = F^{256: 64} (F^{500:256}  (f_g));  \ \  f_d' = F^{256: 64} (F^{1024:256} (f_d)); \nonumber \\
    &f_p' = F^{107: 64} (f_p); \ \
    \text{IF}(f_x, f_y) = F^{64:1} ( F^{128:64}( [f_x', f_y'])); \nonumber 
\end{align}
and late fusion MLP (LF)
\begin{align}
\setlength{\itemsep}{0em}
    &f_x'' = F^{64: 1}(f_x'); \ \ 
    \text{LF}({f_x, f_y}) = F^{2:1} ( [f_x', f_y']), \ \ 
    \nonumber
\end{align}
and the two other methods replace the concatenation step in intermediate fusion (IF) above with
multimodal factorized high-order pooling, MFH~($f_x, f_y$) $ = F_0^{l: 64}([z_0 \ , \ z_0 \ . \ z_1]) )$ where 
$$ z_0 = F_1^{64: l} (f_x') \ . \ F_2^{256: l} (f_y'); \ z_1 = F_3^{64: l} (f_x')\ . \ F_4^{256: l} (f_y'); $$
and block superdiagonal tensor fusion, BLOCK$(f_x, f_y) =$ 
$$
        \text{ReLU} ( (\sum_{r=1}^R D_r \times X_r \times Y_r \times T_r ) \times f_x' \times f_y'), $$
where $F^{d_{\text{in}}:d_{\text{out}}}$ represents a fully connected layer with $d_{\text{in}}$ input channels and $d_{\text{out}}$ output channels followed by a rectified linear unit (ReLU). $f_x$ and $f_y$ can refer to any of the three features $f_g, f_p$ and $f_d$. $[x, y]$ denotes concatenation.
$D_r, X_r, Y_r, T_r$ are the different tensors for the $R$ rank filter decomposition. The default values of $l=1200$ and $R=15$ were used.

The fusion models MFH and BLOCK~\cite{BenYounes_2019_AAAI} are adapted from visual question answering tasks which have a similar setup of imaging and accompanying information (text) and the objective is to combine information effectively.

To train the networks, ranking based loss functions focus on order and are better-suited for the our prediction task compared to regression lossses. Several such loss functions have been proposed, including list-based losses and pairwise losses~\cite{chen2009ranking}. 
In our experiments, we found the following pairwise loss function on features $x$ and event times $y$ to yield the best empirical results:
%




 $\mathcal{L}(x, y) = \sum_{i} \sum_{j: y_j > y_i} \log( 1 + \exp(f(x_j) - f(x_i))).$


\subsection{Experiments and results}
The different models in Section~\ref{nn_methods} are trained with stochastic gradient descent at a learning rate of 0.001 with early stopping and a dropout with probability $p=0.25$. The best L2 weight decay is chosen via hyperparameter tuning. Each setting is run with multiple initializations, and the best performing method and hyperparameter are chosen based on the validation set's C-index metric. Initializations which resulted in early stopping prior to any learning (i.e. number of backpropagation steps < 1000) were discarded.

From the C-index metrics across different folds in Table~\ref{tab:non-linear}~(b), we find that the network finds it difficult to learn good prediction functions for single modalities. This is likely due to overfitting on the small datasets despite early stopping and dropout regularization. The fusion methods sometimes perform better than the single modalities. The time-dependent AUC in Fig.\ref{fig:time_varying_auc}~(c) corresponds to Fold 5. Comparing to Fig.\ref{fig:time_varying_auc}~(b), we note a lower curve for the neural network setting. The VQA fusion techniques, also plotted, are unable to sufficiently improve performance, possibly also due to overfitting due to the large number of parameters that need to be tuned from a small dataset.
\section{Conclusions}


Through experiments on the dataset of 107 patients, we demonstrated the potential of multimodal fusion for recurrence prediction using early and late fusion with Cox proportional hazards model. Our observation that different modalities perform better in different settings highlights the opportunity to improve prediction through multi-modality fusion. We did not, however, find a particular method which outperforms the others across all folds. Neural network methods were found to be harder to generalize when trained on the small dataset.
%

Transfer learning from single-modality datasets could help to better train multimodal fusion networks; this is an interesting direction for future research. 
Multi-stage fusion mimicking biological interactions between imaging and genomics in a structured manner is also of interest in further investigations. 
Additionally, the availability of more paired datasets would be necessary for more rigorous analyses and comparisons. 

\section{Acknowledgment}

The authors would like to thank Deepta Rajan (IBM Research, Almaden) for useful discussions and suggestions.

\clearpage

\bibliographystyle{ieeetr}
\bibliography{references}


\end{document}